\documentclass[twocolumn,prb,aps,superscriptaddress]{revtex4}

\usepackage{graphicx}
\usepackage{dcolumn}
\usepackage{amsmath}

\newlength{\figwidth}
\newlength{\figwidthb}
\begin{document}

\title{X-ray scattering study of pyrochlore iridates: crystal structure, electronic and magnetic excitations}
\author{J. P. Clancy}
\affiliation{Department of Physics, University of Toronto, Toronto, Ontario M5S~1A7, Canada}
\author{H. Gretarsson}
\affiliation{Department of Physics, University of Toronto, Toronto, Ontario M5S~1A7, Canada}
\author{E. K. H. Lee}
\affiliation{Department of Physics, University of Toronto, Toronto, Ontario M5S~1A7, Canada}
\author{Di Tian}
\affiliation{Department of Physics, University of Toronto, Toronto, Ontario M5S~1A7, Canada}
\author{J. Kim}
\affiliation{Advanced Photon Source, Argonne National Laboratory, Argonne, Illinois 60439, USA}
\author{M. H. Upton}
\affiliation{Advanced Photon Source, Argonne National Laboratory, Argonne, Illinois 60439, USA}
\author{D. Casa}
\affiliation{Advanced Photon Source, Argonne National Laboratory, Argonne, Illinois 60439, USA}
\author{T. Gog}
\affiliation{Advanced Photon Source, Argonne National Laboratory, Argonne, Illinois 60439, USA}
\author{Z. Islam}
\affiliation{Advanced Photon Source, Argonne National Laboratory, Argonne, Illinois 60439, USA}
\author{Byung-Gu Jeon}
\affiliation{CeNSCMR, Department of Physics and Astronomy, Seoul National University, Seoul 151-747, Republic of Korea}
\author{Kee Hoon Kim}
\affiliation{CeNSCMR, Department of Physics and Astronomy, Seoul National University, Seoul 151-747, Republic of Korea}
\author{S. Desgreniers}
\affiliation{Laboratoire de physique des solides denses, Department of Physics, University of Ottawa, Ottawa, Ontario, K1N 6N5, Canada}
\author{Yong Baek Kim}
\affiliation{Department of Physics, University of Toronto, Toronto, Ontario M5S~1A7, Canada}
\author{S. J. Julian} 
\affiliation{Department of Physics, University of Toronto, Toronto, Ontario M5S~1A7, Canada}
\author{Young-June Kim}
\email{yjkim@physics.utoronto.ca} 
\affiliation{Department of Physics, University of Toronto, Toronto, Ontario M5S~1A7, Canada}

\date{\today}

\begin{abstract}
We have investigated the structural, electronic, and magnetic properties of the pyrochlore iridates $\rm Eu_2Ir_2O_7$ and $\rm Pr_2Ir_2O_7$ using a combination of resonant elastic x-ray scattering, x-ray powder diffraction, and resonant inelastic x-ray scattering (RIXS).  The structural parameters of $\rm Eu_2Ir_2O_7$ have been examined as a function of temperature and applied pressure, with a particular emphasis on regions of the phase diagram where electronic and magnetic phase transitions have been reported.  We find no evidence of crystal symmetry change over the range of temperatures ($\sim$6 to 300 K) and pressures ($\sim$0.1 to 17 GPa) studied.  We have also investigated the electronic and magnetic excitations in single crystal samples of $\rm Eu_2Ir_2O_7$ and $\rm Pr_2Ir_2O_7$ using high resolution Ir L$_3$-edge RIXS.  In spite of very different ground state properties, we find these materials exhibit qualitatively similar excitation spectra, with crystal field excitations at $\sim$3-5 eV, spin-orbit excitations at $\sim$0.5-1~eV, and broad low-lying excitations below $\sim$0.15 eV. In $\rm Eu_2Ir_2O_7$ we observe highly damped magnetic excitations at $\sim$45~meV, which display significant momentum dependence.  We compare these results with recent dynamical structure factor calculations.
\end{abstract}

 \maketitle

\section{Introduction}

The pyrochlore iridates $\rm R_2Ir_2O_7$ (R = Y or rare-earth lanthanides) possess a unique combination of extended 5d orbitals, strong electronic correlations, small magnetic moments, geometric frustration, and large spin-orbit coupling effects.  In the case of R = Pr, these features appear to give rise to a novel metallic spin liquid ground state and an anomalous Hall effect\cite{Nakatsuji2006,Machida2007,Machida2010,SBLee2013}.  In the case of R = Eu, Y, Nd, Sm, and Lu, these features have been predicted to give rise to new topological phases, such as the Weyl semi-metal state\cite{Wan2011,Balents-Physics,Witczak2011}.  Unlike a topological insulator, which has the bulk electronic properties of an insulator with topologically protected conducting states on the surface, the Weyl semi-metal has both surface and bulk states at the Fermi surface. The low energy bulk band structure is described by a Dirac quasiparticle-like linear dispersion, while the surface states are characterized by a Fermi arc. Furthermore, while topological insulators require time-reversal symmetry, the Weyl semi-metal is realized in three-dimensional magnetic solids which break time-reversal symmetry. Therefore, a significant amount of theoretical work has gone into determining the magnetic ground state of the $\rm R_2Ir_2O_7$ systems\cite{Wan2011,Witczak2011,Maiti2009,Yang2010}.  In particular, the Weyl semi-metal phase in $\rm Y_2Ir_2O_7$ has been predicted to display an all-in/all-out spin configuration\cite{Wan2011}.  In the all-in/all-out structure, the spins at the four corners of each Ir-tetrahedron point directly inwards (all-in) or outwards (all-out) along the local $\langle$111$\rangle$ direction (i.e. towards or away from the center of the tetrahedron), while the spins on each neighboring tetrahedron align in the opposite direction. A detailed theoretical account of Weyl semi-metal physics and other topological phases is provided in recent review articles \cite{Witczak2014,Vafek2014}.

Experimentally, bulk characterization measurements\cite{Taira2001,Yanagishima2001,Matsuhira2007,Matsuhira2011,Zhao2011,Ishikawa2012} on the R$_2$Ir$_2$O$_7$ series provide evidence of a magnetic phase transition at temperatures ranging from $T_c$ = 36~K (R = Nd) to 150~K (R = Y).  In the case of R = Nd, Sm, and Eu, this magnetic transition is also accompanied by an electronic transition from metal to insulator.  Magnetic susceptibility measurements on $\rm R_2Ir_2O_7$ (R = Y, Sm, Eu, and Lu) reveal a small anomaly at $T_c$ and significant hysteresis between field-cooled and zero-field-cooled curves \cite{Taira2001,Matsuhira2007,Matsuhira2011,Ishikawa2012}. This has been interpreted as evidence of either a spin glass \cite{Taira2001} or a more complicated antiferromagnetic state \cite{Matsuhira2007}. Muon spin rotation ($\mu$SR) measurements performed on $\rm Eu_2Ir_2O_7$ by Zhao {\it et al.} suggest the presence of commensurate long-range order below $T_c = 120$~K \cite{Zhao2011}.  A similar result has also been reported for Yb$_2$Ir$_2$O$_7$ with $T_c$ = 130 K\cite{Disseler2012a}.  However, $\mu$SR studies on $\rm Nd_2Ir_2O_7$, Sm$_2$Ir$_2$O$_7$, and Y$_2$Ir$_2$O$_7$ indicate that only short-range (i.e. spin-glass-like) magnetic order develops in the hysteretic region below $T_c$, with true long-range order only forming at much lower temperatures\cite{Disseler2012b, Graf2014, Disseler2012a}.  In a recent analysis by Disseler it is argued that $\mu$SR data on $\rm Y_2Ir_2O_7$ and $\rm Eu_2Ir_2O_7$ is only consistent with an all-in/all-out magnetic structure\cite{Disseler2014}.

Although bulk susceptibility and $\mu$SR studies have clearly established the presence of broken time-reversal symmetry in these compounds, very little direct information regarding the ordered moment direction and periodicity is available to test theoretical predictions. Since Ir is a strong neutron absorber, studying iridate compounds with neutron scattering is technically challenging. Nevertheless, several attempts have been made to study magnetic order in pyrochlore iridates using neutron powder diffraction. Disseler {\it et al.} were  unable to detect a magnetic signal in their neutron powder diffraction study of $\rm Nd_2Ir_2O_7$\cite{Disseler2012b}. Similarly, Shapiro and coworkers\cite{Shapiro2012} did not detect ordering of Ir moments in their study of $\rm Y_2Ir_2O_7$. However, the latter measurements did establish an upper limit on the size of the possible Ir ordered moment, which was set at 0.2 $\mu_B$ (for ${\bf q} \neq 0$) or 0.5 $\mu_B$ (for {\bf q} = 0). This result was subsequently confirmed by Disseler {\it et al.}\cite{Disseler2012a}

Indirect confirmation of the all-in/all-out structure has been provided by two recent neutron and x-ray diffraction experiments. In their neutron study of $\rm Nd_2Ir_2O_7$, Tomiyasu {\it et al.}\cite{Tomiyasu2012} were able to detect magnetic intensity at low temperatures due to the ordering of Nd moments, even though Ir ordering was below the experimental detection limit. Nd moments were found to order in an all-in/all-out fashion, leading the authors to conclude that Ir moments were likely to adopt the same configuration. In a recent resonant x-ray diffraction experiment on Eu$_2$Ir$_2$O$_7$, Sagayama {\it et al.}\cite{Sagayama2013} directly observed the ordering of Ir magnetic moments with a wave vector of {\bf q} = 0. Unfortunately, this result alone does not provide sufficient information to unambiguously determine the spin structure. However, Sagayama and coworkers were able to argue that the all-in/all-out structure is the only symmetry allowed spin structure which does not result in a structural distortion, and showed that the crystal symmetry remains unchanged across $T_c$ \cite{Sagayama2013}. This lack of structural symmetry change at $T_c$ is supported by recent high-resolution x-ray powder diffraction measurements\cite{Takatsu2014}. However, these results are at odds with previous Raman scattering measurements on R = Eu and Sm pyrochlores, which report the appearance of new peaks below $T_c$ \cite{Hasegawa2010}.

The physics of pyrochlore iridates can also be explored by applying hydrostatic pressure. High pressure transport studies have revealed new electronic phases in both $\rm Nd_2Ir_2O_7$ \cite{Sakata2011} and $\rm Eu_2Ir_2O_7$ \cite{Tafti2012}. In the case of Eu$_2$Ir$_2$O$_7$, Tafti {\it et al.} discovered a pressure-induced electronic transition at $\sim$6 GPa.  At low temperature (T $<$ $T_c$), applied pressure drives Eu$_2$Ir$_2$O$_7$ from an insulator to an unconventional metal ($\partial \rho$/$\partial T <$ 0 and non-Fermi-liquid-like temperature dependence).  At high temperature (T $>$ $T_c$), applied pressure drives the system from an incoherent metal (high resistivity, power law temperature dependence, and $\partial \rho$/$\partial T <$ 0) to a conventional metallic state ($\partial \rho$/$\partial T >$ 0).  The value of $T_c$ for Eu$_2$Ir$_2$O$_7$ was found to exhibit very little pressure dependence up to $\sim$12 GPa. In the case of Nd$_2$Ir$_2$O$_7$, Sakata {\it et al.} found that the metal-insulator transition is gradually suppressed by applied pressure, with the insulating state disappearing above $\sim$10~GPa. A new low temperature transition is observed in the pressure-induced metallic state, accompanied by a drop in resistivity and potential ferromagnetic (2-in/2-out) ordering of Nd moments.  Whether the structure of these materials stays the same in these high pressure phases remains unknown. A careful investigation of structural symmetry is therefore extremely important for our understanding of these new pressure-induced electronic states.

Even in the absence of a distinct transition or crystal symmetry change, variations in structural parameters appear to play a key role in determining the electronic and magnetic properties of the pyrochlore iridates. In particular, previous experimental work has shown that both of these properties are highly sensitive to the A-site cation size \cite{Taira2001,Yanagishima2001}. As shown in Fig.~1, the pyrochlore crystal structure is described by space group $Fd\bar{3}m$ (\#227), with A-site atoms at the 16d position (0.5, 0.5, 0.5), Ir at the 16c position (0, 0, 0), O1 at 48f ($x$, 0.125, 0.125), and O2 at 8a (0.375, 0.375, 0.375). Only two free parameters in this structure -- the lattice constant $a$ and the $x$-coordinate associated with the O1 site -- are adjustable. All pyrochlore iridates have $x$ larger than the value for an ideal IrO$_6$ octahedra ($x_c = 5/16$), indicating that a compressive trigonal distortion is present. Larger A-site ions tend to result in a smaller distortion, which is believed to be responsible for increased hopping and a stronger metallic character\cite{Witczak2014}.

\begin{figure}
\begin{center}
\includegraphics[angle=0,width=3.1in]{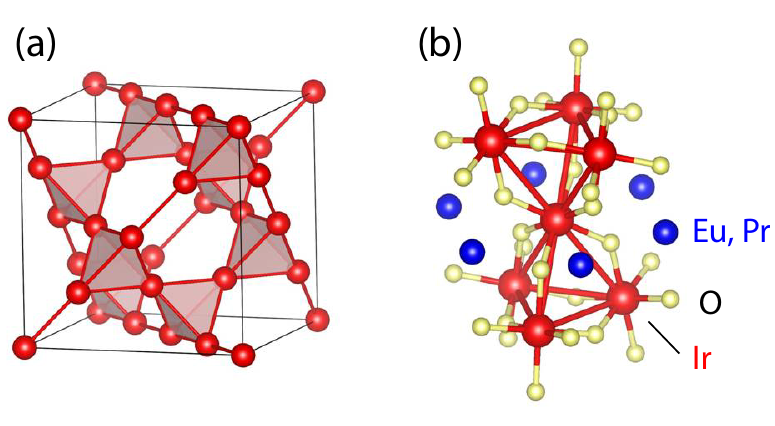}
\end{center}
\caption{(Color online) Crystal structure of the pyrochlore iridates Eu$_2$Ir$_2$O$_7$ and Pr$_2$Ir$_2$O$_7$. (a) The Ir sublattice, which forms a network of corner-sharing tetrahedra. (b) The local environment at the Ir site.  Each Ir atom lies at the center of a trigonally compressed IrO$_6$ octahedra, surrounded by a hexagonal ring of Eu/Pr atoms.}\label{fig1}
\end{figure}

The examples above illustrate the importance of structural details in determining the electronic and magnetic properties of pyrochlore iridates. In the first half of this article, we report comprehensive x-ray diffraction measurements which address several of the unresolved questions raised above. First, we present a resonant x-ray diffraction study of temperature dependence in single crystal $\rm Eu_2Ir_2O_7$. We observe x-ray diffraction peaks which violate structure factor extinction rules when x-ray energy is tuned near the Ir L$_3$ absorption edge. These peaks arise from so-called Templeton scattering, which reflects the anisotropic charge distribution near the iridium atoms and is very sensitive to local structural distortions. We observe no significant changes in peak intensity as the sample passes through the magnetic transition at $T_c$, supporting the idea that the magnetic transition in Eu$_2$Ir$_2$O$_7$ occurs without structural distortion. We also present an x-ray powder diffraction study of $\rm Eu_2Ir_2O_7$ as a function of applied pressure up to $\sim$17 GPa. We observe no evidence of structural transitions induced by applied pressure, and find that all structural parameters evolve monotonically over the range of pressures studied. Therefore, our results suggest that the pressure-induced metal-insulator transition observed by Tafti {\it et al.}\cite{Tafti2012} is of purely electronic origin.

We were unable to detect magnetic Bragg peaks associated with magnetic order in our resonant x-ray diffraction studies. However, the magnetic ordering wave vector, {\bf q}, can often be revealed by studying the momentum dependence of magnetic excitations. We have thus carried out a Resonant Inelastic X-ray Scattering (RIXS) investigation to study the magnetic excitation spectrum of single crystal $\rm Eu_2Ir_2O_7$, which will be discussed in the second half of this article. RIXS is a second-order scattering process which can be used to probe elementary excitations involving spin, orbital, charge, and lattice degrees of freedom \cite{Ament2011a,Ament2011b}. RIXS has made particularly significant contributions to our understanding of the physics of iridates \cite{JKim2012a,JKim2012b,JKim2014,Gretarsson2013a,Gretarsson2013b,Liu2012,Yin2013,Moretti2014,Lupascu2014}. Among its many advantages, RIXS only requires very small sample volumes, and is unaffected by the neutron absorption issues which hinder inelastic neutron studies. The {\it d-d} excitations (or spin-orbit excitations) of $\rm Eu_2Ir_2O_7$ and $\rm Y_2Ir_2O_7$ have been studied in earlier work by combined RIXS and theoretical quantum chemistry calculations \cite{Hozoi2014}.

In this article, we report low energy magnetic excitations in single crystal $\rm Eu_2Ir_2O_7$. Unlike the insulating iridates $\rm Sr_2IrO_4$ and Sr$_3$Ir$_2$O$_7$, in which well-defined magnon modes have been observed\cite{JKim2012a,JKim2012b}, here we find only a highly damped excitation, with weak momentum dependence and a very broad peak width. This result is found to be consistent with theoretical calculations of the dynamic structure factor for all-in/all-out magnetic order in the intermediate-coupling regime.  RIXS measurements on single crystal $\rm Pr_2Ir_2O_7$ reveal a very similar excitation spectrum, distinguished by larger non-cubic crystal field splitting and a broad, low-lying feature which may be attributed to paramagnetic fluctuations.

\section{Experimental methods and sample characterization}

For resonant elastic and inelastic scattering experiments, single crystal samples of $\rm Eu_2Ir_2O_7$ and $\rm Pr_2Ir_2O_7$ were grown using KF flux methods. As noted in Ref.~\onlinecite{Hozoi2014}, resistivity measurements performed on the resulting $\rm Eu_2Ir_2O_7$ single crystal revealed almost metallic behavior, with a residual resistivity ratio of $\rho_{4K}$/$\rho_{300K}$ $\sim$0.7. Issues with sample stoichiometry are known to be very common in this family of materials\cite{Ishikawa2012}, and this result was initially interpreted as evidence of a slight excess of Ir. However, subsequent characterization via electron probe microanalysis (EPMA) has revealed that the actual sample composition is Eu-rich: Eu$_{2(1-x)}$Ir$_{2(1+x)}$O$_{7+\delta}$ with x = -0.09(2) and $\delta$ = 0.06(2). A thorough investigation of disorder and sample stoichiometry effects on the transport properties of $\rm Eu_2Ir_2O_7$ is provided in Ref.~\onlinecite{Ishikawa2012}. The Eu-rich region of the Eu$_{2(1-x)}$Ir$_{2(1+x)}$O$_{7+\delta}$ phase diagram has received comparatively little attention.  In our sample, we find that x = -0.09 is sufficient to split the transition at $T_c$ $\sim$ 120 K into separate magnetic ($T_M$ $\sim$ 155 K) and electronic ($T_{E}$ $\sim$ 60 K) phase transitions. Electrical resistivity and magnetic susceptibility data for this sample can be found in the supplemental material accompanying this article. A comparison with recent measurements on ``near-stoichiometric'' Eu$_2$Ir$_2$O$_7$ (x $<$ 0.05)\cite{Uematsu2015} shows that the high energy RIXS spectrum exhibits very little stoichiometry dependence (i.e. the {\it d-d} excitations are essentially unchanged). 

For x-ray powder diffraction experiments, a polycrystalline sample of $\rm Eu_2Ir_2O_7$ was synthesized using standard solid-state reaction methods. A mixture of $\rm Eu_2O_3$ and $\rm IrO_2$ with purity of 99.99\% was ground in a stoichiometric molar ratio, pelletized, and then heated in air at 1000 $^\circ$C for 100 hours. The resulting material was reground, pressed into pellets, and resintered at the same temperature for an additional 150 hours, with two intermediate regrindings. The phase purity of the sample was verified by x-ray powder diffraction, and the sample stoichiometry was determined from Rietveld refinements. The polycrystalline sample was found to be slightly Ir-rich, with a composition of Eu$_{1.97}$Ir$_{2.03}$O$_{7}$ (x = 0.015(6)).   

Resonant x-ray scattering measurements were performed on single crystal $\rm Eu_2Ir_2O_7$ using beamline 6-ID-B at the Advanced Photon Source. The incident x-ray energy was tuned near the Ir L$_3$ absorption edge at $\sim$11.22 keV. Incident photons were linearly polarized perpendicular to the vertical scattering plane ($\sigma$ polarization). Resonant magnetic scattering rotates the plane of linear polarization into the scattering plane ($\pi$ polarization). In contrast, charge scattering does not change the polarization of the scattered photons. As a result, polarization analysis of the scattered beam can be used to distinguish the magnetic ($\sigma-\pi$) and charge ($\sigma-\sigma$) scattering contributions. The (3,3,3) reflection from single crystal aluminum was used as a polarization and energy analyzer ($2\theta_a$ $\sim$ 90.3$^{\circ}$). The sample was mounted on the coldfinger of a closed-cycle refrigerator capable of reaching temperatures from 6 K to 300 K. Single crystal Eu$_2$Ir$_2$O$_7$ grows with facets along the $\langle$111$\rangle$ direction, and our measurements primarily focused on reflections along or close to the surface normal direction.  A series of measurements were also carried out on a second sample with a cut and polished $\langle$100$\rangle$ surface. The results from the $\langle$100$\rangle$ and $\langle$111$\rangle$ surfaces are qualitatively similar, hence only the $\langle$111$\rangle$ results will be presented here.

High-pressure x-ray diffraction measurements were performed using the Hard X-ray Micro-Analysis (HXMA) 06ID-1 beamline at the Canadian Light Source. Finely ground Eu$_2$Ir$_2$O$_7$ was loaded into a diamond anvil cell, and measured under applied pressures of up to $\sim$17 GPa. The pressure was tuned with a precision of $\pm 0.2$~GPa using the R1 fluorescent line from a ruby chip placed inside the sample space. Data sets were collected using two different choices of quasistatic pressure transmitting medium: (1) silicone fluid (polydimethylsiloxane, 1 cSt) and (2) a methanol-ethanol-water (MEW) mixture in a volume ratio of 16:3:1. X-ray diffraction data were collected using angle-dispersive techniques, employing high-energy x rays ($E_i$ = 24.350 keV) and a MAR345 image plate detector. Additional details of the experimental setup are reported elsewhere\cite{Smith2009}. Structural parameters were extracted from full profile Rietveld refinements carried out using the GSAS software package\cite{GSAS}.

RIXS measurements were performed on single crystal $\rm Eu_2Ir_2O_7$ and $\rm Pr_2Ir_2O_7$ using the MERIX spectrometer on beamline 30-ID-B at the Advanced Photon Source. A double-bounce diamond-(1,1,1) primary monochromator, a channel-cut Si-(8,4,4) secondary monochromator, and a spherical (2 m radius) diced Si-(8,4,4) analyzer crystal were used to obtain an overall energy resolution of 35 meV (full width at half maximum [FWHM]).  In order to minimize the elastic background intensity, measurements were carried out in horizontal scattering geometry, focusing on wave vectors in the vicinity of {\bf Q} = (7.5, 7.5, 7.5) for which the scattering angle $2 \theta$ is close to 90 degrees.

\section{Experimental results: structure}

\subsection{Resonant x-ray diffraction}

The pyrochlore $\rm Eu_2Ir_2O_7$ crystallizes in the cubic space group $Fd\bar{3}m$ with a lattice parameter of $a = 10.274(3)$ \AA.\cite{Millican2007} In each unit cell there are 8 formula units, meaning that the theoretically predicted all-in/all-out magnetic order will give rise to {\bf q} = 0 magnetic Bragg peaks. In addition, there are a number of other non-collinear antiferromagnetic structures, predicted to be similar in energy to the all-in/all-out state\cite{Witczak2011}, which also correspond to {\bf q} = 0 type order. In general, {\bf q} = 0 magnetic order is very difficult to detect due to the large scattering contribution from structural Bragg peaks that coincide with the magnetic peaks. However, because the atoms in $\rm Eu_2Ir_2O_7$ exclusively occupy special high symmetry positions in the pyrochlore lattice, there is an extinction rule which governs reflections with $h, k = 4n$, $l = 4n + 2$; i.e. structural peaks such as (4,4,2) and (4,4,6) are absent. Therefore, {\bf q} = 0 magnetic peaks may be observable in these positions. Similarly, there are very weak structural Bragg peaks, such as (5,5,5), which might allow one to study magnetic peak intensity superimposed on the structural intensity.

\begin{figure}
\centering
\includegraphics[angle=0,width=3.1in]{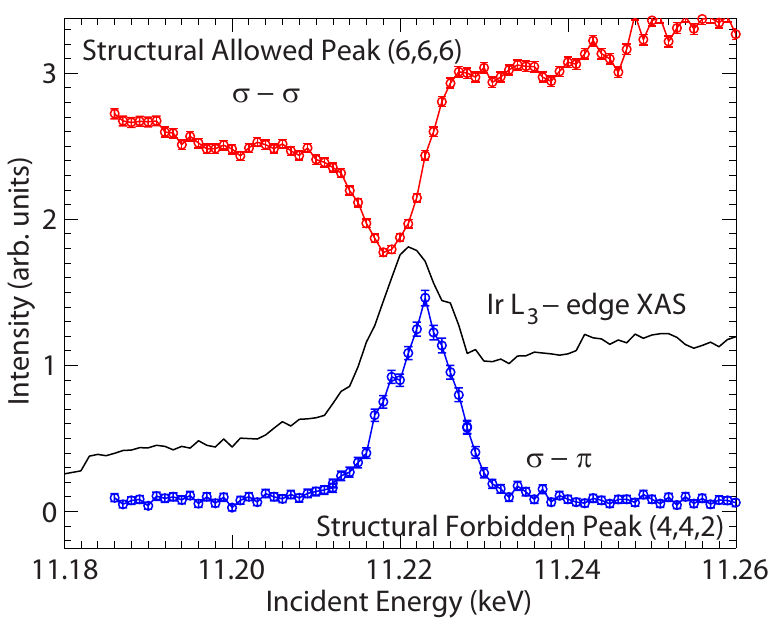}
\caption{(Color online) Resonant x-ray scattering intensity as a function of incident photon energy in $\rm Eu_2Ir_2O_7$.  Note the contrasting behavior of a normal structural peak, (6,6,6) [shown in red], and the resonantly enhanced intensity of a forbidden reflection, (4,4,2) [shown in blue].  These scans were collected using a polarization analyzer in $\sigma-\sigma$ and $\sigma-\pi$ configurations, respectively (see main text for details).  The x-ray absorption spectrum at the Ir L$_3$-edge is illustrated by a solid black line.}\label{fig2}
\end{figure}

We concentrated our attention on two of these forbidden or weak Bragg peak positions, (4,4,2) and (5,5,5), and investigated the energy, momentum, polarization, and temperature dependence of these peaks. The energy dependence of the (4,4,2) peak is shown in Fig.~\ref{fig2}. A clear resonant enhancement is observed at this position near E = 11.223 keV. This resonance behavior is contrasted with the energy dependence of a regular structural Bragg peak, (6,6,6), which exhibits typical intensity modulation near the absorption edge. The Ir L$_3$ absorption profile, obtained by monitoring x-ray fluorescence from the sample, is plotted to illustrate that the (6,6,6) peak intensity is inversely correlated with the absorption coefficient. We note that the resonant enhancement for the (4,4,2) peak occurs at an energy slightly above the XAS maximum (E = 11.221 keV). This resonance behavior is quite different from that of the magnetic peaks reported for other iridates such as Sr$_2$IrO$_4$\cite{BJKim2009}, Na$_2$IrO$_3$\cite{Liu2011}, and Sr$_3$Ir$_2$O$_7$\cite{Boseggia2012}, all of which display a resonance several eV {\it below} the XAS maximum.

In spite of the fact that the (4,4,2) peak displays significant intensity in the $\sigma-\pi$ scattering channel, this difference in photon energy dependence provides strong evidence against assigning this peak as a magnetic reflection. The lack of temperature dependence for this peak (discussed below) further supports this interpretation. Instead, we attribute the resonant intensity at these forbidden peak positions to Templeton scattering, or the Anisotropic Tensor of Susceptibility (ATS). This scattering arises from the anisotropic nature of the d orbitals and the shape of the local charge distribution around the Iridium site. In previous studies, ATS has been used to probe the $j_{eff}$ = 1/2 character of electronic wavefunctions in CaIrO$_3$\cite{Ohgushi2013} and Eu$_2$Ir$_2$O$_7$\cite{Uematsu2015}. Here, we can use the structural sensitivity of ATS to investigate the degree of distortion around the Ir atoms.  

\begin{figure}
\begin{center}
\includegraphics[angle=0,width=3.1in]{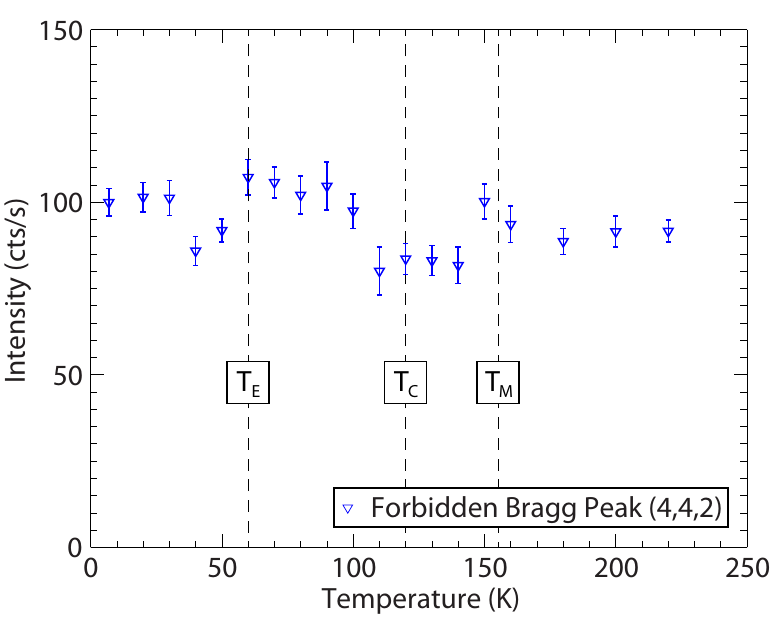}
\end{center}
\caption{(Color online) Temperature dependence of the resonantly enhanced (4,4,2) peak in $\rm Eu_2Ir_2O_7$ (See Fig.~2).  Note the absence of any anomalies associated with the magnetic and electronic phase transitions.}\label{fig3}
\end{figure}

The temperature dependence of the structurally forbidden (4,4,2) peak is shown in Fig.~\ref{fig3}. Note that, within experimental uncertainty, the peak intensity does not change from room temperature down to 6 K. In particular, we do not observe any intensity changes associated with the magnetic or electronic transition temperatures: $T_c \sim$ 120 K (for near-stoichiometric Eu$_2$Ir$_2$O$_7$), $T_M \sim$ 155 K and $T_E \sim$ 60 K (for Eu-rich Eu$_2$Ir$_2$O$_7$). This lack of temperature dependence indicates (1) the observed peak is not magnetic in origin, and (2) there are no local structural distortions at $T_c$, $T_M$, or $T_E$. In order to verify that these observations are general beyond the (4,4,2) reflection, similar measurements were also performed for the (5,5,5) Bragg reflection. (5,5,5) is an allowed Bragg peak, but one with a very small structure factor (over 10$^3$ times weaker than the nearby (4,4,4) reflection). As in the case of (4,4,2), we observe resonantly enhanced intensity in the $\sigma-\pi$ channel, with no apparent temperature dependence. We also scanned along several high symmetry directions in reciprocal space in order to search for {\bf q} $\neq$ 0 magnetic order. No significant signal above background was observed in any of these scans.

\subsection{High pressure crystal structure}

\begin{figure}
\begin{center}
\includegraphics[angle=0,width=3.1in]{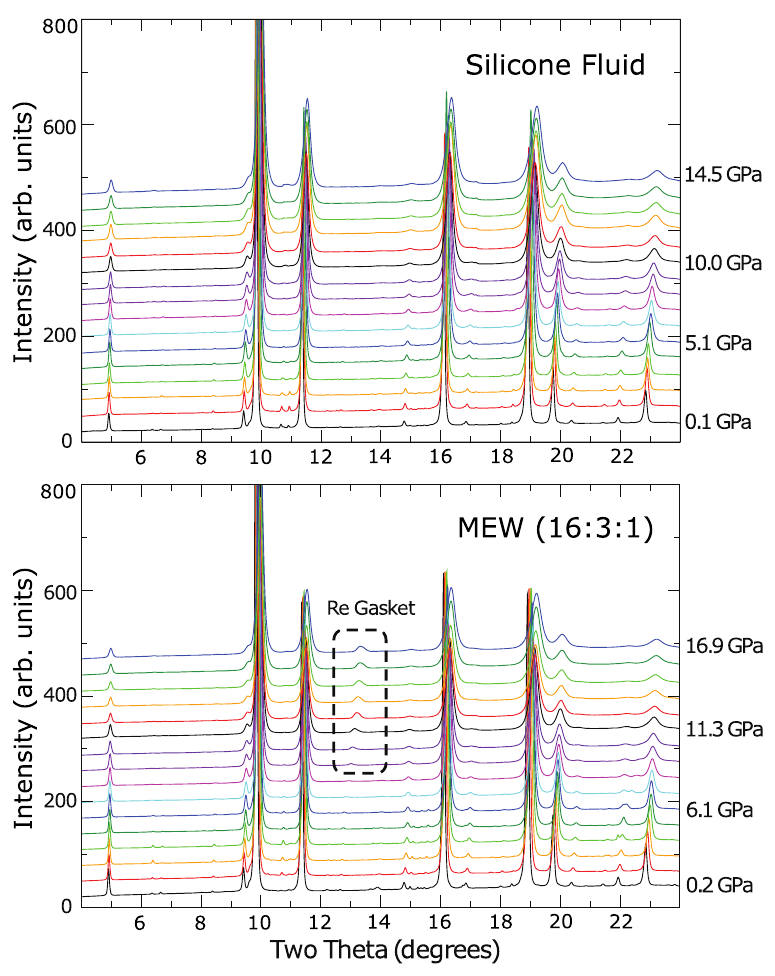}
\end{center}
\caption{(Color online) X-ray diffraction patterns for $\rm Eu_2Ir_2O_7$ as a function of applied pressure.  Patterns have been vertically offset for clarity, and are arranged in order of increasing pressure from bottom to top.  Patterns in the upper (lower) panel were obtained using silicone fluid (methanol-ethanol-water mixture [MEW]) as pressure transmitting medium.  Representative patterns shown here correspond to P = 0.1, 0.9, 1.8, 2.6, 3.8, 5.1, 6.1, 6.9, 7.9, 8.9, 10.0, 11.0, 12.0, 12.8, 13.6, 14.5 GPa for silicone fluid, and P = 0.2, 1.7, 2.9, 4.2, 5.3, 6.1, 7.0, 8.1, 9.1, 10.2, 11.3, 12.6, 13.5, 14.7, 15.8, 16.9 GPa for MEW.  All patterns were collected at room temperature. }\label{fig4}
\end{figure}

In Fig.~\ref{fig4}, we show the evolution of the x-ray diffraction pattern for $\rm Eu_2Ir_2O_7$ as a function of applied pressure. The data in Fig.~\ref{fig4}(a) were obtained using silicone fluid as a pressure transmitting medium, whereas the data in Fig.~4(b) were obtained using MEW.  The diffraction patterns from the two pressure transmitting media are almost identical, save for the small peak at 2$\theta$ $\sim$ 13$^{\circ}$ which appears in Fig.~4(b) due to scattering from the Re gasket. These diffraction patterns reveal a gradual contraction of the lattice with increasing pressure, as all Bragg peaks move towards progressively higher $2 \theta$ angles. The diffraction peaks also become noticeably broader at higher pressures, a common effect which arises due to strain-broadening. However, there is no evidence of new peaks forming, or of existing peaks splitting, as the pressure is varied. These results indicate that no structural phase transitions occur in Eu$_2$Ir$_2$O$_7$ up to P $\sim$ 17 GPa. 

It is important to be cautious about experimentally realizing pure hydrostatic pressure conditions. This is particularly true during low temperature experiments, where there is a strong possibility that pressure transmitting media may solidify and give rise to anisotropic pressure conditions. However, at room temperature, there are many liquid pressure media which work well to preserve hydrostatic conditions up to relatively high pressures. In this experiment we have used two different choices of pressure transmitting medium -- MEW (16:3:1) and silicone fluid. The hydrostatic limit for MEW has been found to be $\sim$10.5 GPa\cite{Klotz2009}, while the reported limits for silicone fluid range from $\sim$3 GPa\cite{Klotz2009} to over 10 GPa\cite{Ragan1996, Shen2004}, depending on specific chemical composition and viscosity.  The low viscosity silicone fluid used in this experiment is expected to offer similar, if not slightly better, performance than MEW over the measured pressure range.  Most importantly, we note that the results from both media agree within error bars, giving credence to our experimental observations.

The diffraction patterns were analyzed by Rietveld refinements carried out using the GSAS software package\cite{GSAS}. The lattice parameters, atomic positions, thermal parameters, lineshape, and background were all refined in this process. The diffraction patterns were well-described by a $Fd\bar{3}m$ structural model from ambient pressure up to P = 14.5 GPa for data obtained with the silicone fluid pressure medium, and from P = 0.2 GPa to P = 16.9 GPa for data obtained with MEW. Goodness-of-fit parameters, $R_{wp}$ and $R_p$, indicate that the quality of the refinements remains good across the full pressure range. This provides further evidence against the occurrence of any pressure-induced structural distortions. Additional details of the structural refinements can be found in the supplemental material accompanying this article.

\begin{figure}
\begin{center}
\includegraphics[angle=0,width=3.1in]{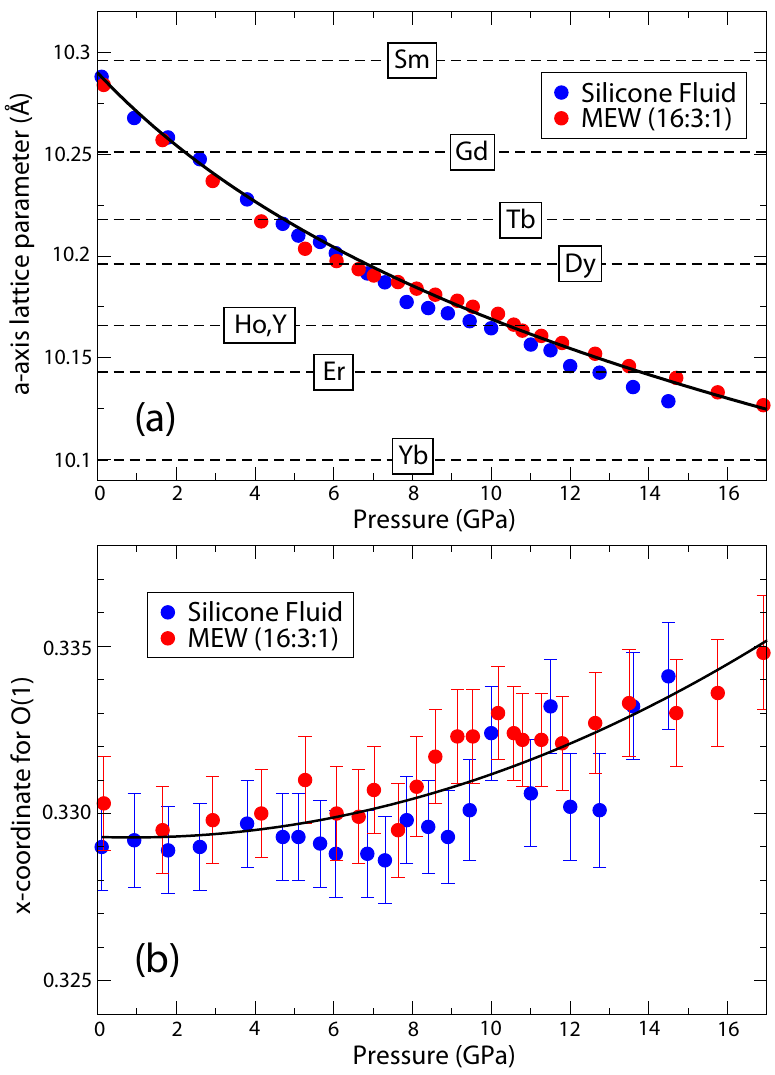}
\end{center}
\caption{(Color online) Pressure dependence of structural parameters for Eu$_2$Ir$_2$O$_7$.  These parameters were obtained from Rietveld refinements performed on the data from Fig.~4.  There are only two free parameters in the pyrochlore crystal structure: (a) the cubic lattice constant $a$, and (b) the $x$-coordinate of O(1), the oxygen atom which sits at the 48f site.  The solid line in (a) is derived from a fit to the Murnaghan equation, while the solid line in (b) is a simple polynomial guide-to-the-eye.  For comparison, the ambient pressure lattice constants for various other pyrochlore iridates $\rm R_2Ir_2O_7$ have been marked by dashed horizontal lines.  The lattice constant for R = Pr is off-scale at {\it a} = 10.396 {\AA}.  The $x$-coordinate for an undistorted IrO$_6$ octahedron is $x = 5/16 = 0.3125$.}\label{fig5}
\end{figure}

In Fig.~\ref{fig5}(a), the refined lattice parameter is plotted as a function of applied pressure for both choices of pressure transmitting media. Both data sets show that the cubic lattice parameter evolves monotonically as a function of pressure. The full pressure range can be well fit by the Murnaghan equation, which describes the relation between hydrostatic pressure and volume contraction. This equation can be written as: 

\begin{equation} 
P(V) = \frac{K_0}{K_0'}\bigg[\Big(\frac{V}{V_0}\Big)^{-K_0'}-1\bigg],
\end{equation}

\noindent where $K_0$ represents the modulus of incompressibility at ambient pressure, $K_{0}^{\prime}$ represents the first derivative of $K_0$ with respect to pressure, and $V_0$ represents the volume of the unit cell at ambient pressure.  From the fit provided in Fig. 5(a) we can extract values of $K_{0}$ = 166(4) GPa and $K_{0}^{\prime}$ = 27.8(5) for Eu$_2$Ir$_2$O$_7$.  We note that the value of $K_{0}^{\prime}$ is surprisingly large, indicating that Eu$_2$Ir$_2$O$_7$ stiffens rapidly as a function of applied pressure.  This result holds true whether the data is fit to Murnaghan or Birch-Murnaghan equations of state.

Small changes in slope can be observed in the intermediate pressure range, at $\sim$7 GPa (silicone fluid) and 10 GPa (MEW), but these changes appear to be well within the experimental uncertainties. The difference between the silicone fluid and MEW data sets grows progressively larger with increasing pressure, which is likely due to deviations from ideal hydrostatic conditions. For comparison, the ambient pressure lattice constants for pyrochlore iridates with smaller R-site ions such as Gd, Tb, Dy, Ho, Y, Er, and Yb, have also been marked in Fig. 5(a). This allows us to directly examine the relation between chemical pressure (tuned by R-site substitution) and actual hydrostatic pressure.  The substitution of Eu$^{3+}$ for Er$^{3+}$ for example, which corresponds to $\sim$6\% reduction in ionic radius, is found to be equivalent to an applied pressure of $\sim$14 GPa.  

Apart from the cubic lattice parameter, there is only one free structural parameter in the pyrochlore lattice: the $x$-coordinate associated with the O(1) site. This positional parameter provides a measure of the distortion of the IrO$_6$ octahedra, with $x = 5/16 = 0.3125$ corresponding to ideal, undistorted octahedra. For $x$ $>$ 5/16 the IrO$_6$ octahedra become trigonally compressed, while for $x$ $<$ 5/16 they become trigonally elongated.  In Fig.~\ref{fig5}(b) we plot the evolution of $x$ as a function of pressure. A small, but finite, increase in $x$ is clearly observed, suggesting that the IrO$_6$ octahedra become increasingly more distorted at higher pressures. The degree of distortion, $\Delta x \equiv x-0.3125$, is $\sim$25-30\% larger at $P \sim 17$~GPa than it is at ambient pressure.

\section{Experimental results: excitations}

\begin{figure}
\begin{center}
\includegraphics[angle=0,width=3.1in]{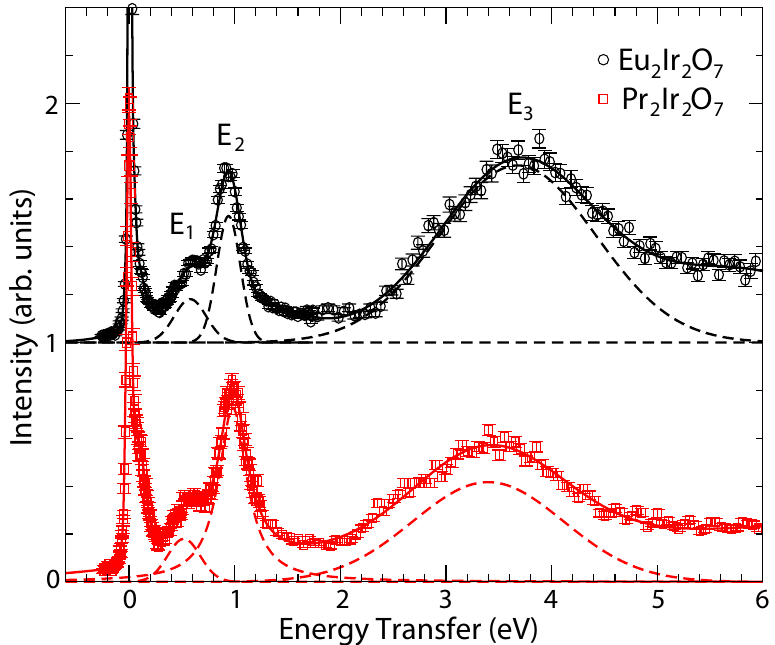}
\end{center}
\caption{(Color online) Resonant inelastic x-ray scattering (RIXS) scans performed on $\rm Eu_2Ir_2O_7$ and $\rm Pr_2Ir_2O_7$ single crystal samples.  Both scans were carried out at room temperature using an incident photon energy of 11.217 keV (near the Ir L$_3$-edge).  The momentum transfer corresponds to {\bf Q} = (7.5, 7.5, 7.5), an $L$-point on the Brilloiun zone boundary. }\label{fig6}
\end{figure}

The RIXS process at the L$_3$ edge of Ir (or any other d electron system) is a second order process consisting of two dipole transitions ($2p \rightarrow 5d$ followed by $5d \rightarrow 2p$). In Fig.~\ref{fig6}, we show representative energy scans obtained from RIXS measurements on $\rm Eu_2Ir_2O_7$ and Pr$_2$Ir$_2$O$_7$ single crystals. These scans contain several prominent features. First, there is a sharp elastic peak at zero energy loss, which provides a measure of the experimental energy resolution ($\sim$35 meV FWHM). This is followed by three strong inelastic features: two sharp low energy excitations around 0.6-1 eV ($E_1$ and $E_2$) and an extremely broad high energy excitation around 2.5-5 eV ($E_3$).  Quantum chemistry calculations\cite{Hozoi2014} have shown that $E_1$ and $E_2$ arise from {\it d-d} excitations within the t$_{2g}$ manifold, while $E_3$ arises from transitions between the t$_{2g}$ and e$_g$ bands. We will demonstrate that there is also a fourth inelastic feature, which can be observed as a small shoulder on the right-hand side of the elastic peak.  This low-lying feature is particularly prominent in Eu$_2$Ir$_2$O$_7$, where we will argue that it must be magnetic in origin.

Before examining the detailed momentum and temperature dependence of the RIXS spectrum for Eu$_2$Ir$_2$O$_7$, we first offer some general comparisons between Eu$_2$Ir$_2$O$_7$ and $\rm Pr_2Ir_2O_7$.  The R = Pr compound appears to be unique among rare-earth pyrochlore iridates, in that it is the only (intrinsically) metallic compound. As one can see from the data in Fig.~\ref{fig6}, the RIXS spectrum of Pr$_2$Ir$_2$O$_7$ is qualitatively very similar to that of $\rm Eu_2Ir_2O_7$. This similarity is expected for the {\it d-d} excitations, as the energies of $E_1$, $E_2$, and $E_3$ are largely determined by local structure (Pr$_2$Ir$_2$O$_7$ shares the same $Fd\bar{3}m$ crystal structure with a slightly expanded lattice parameter, $a=10.41$~\AA). However, the existence of a broad, low-lying excitation is quite interesting, since no magnetic order has been observed in the R = Pr compound down to 70 mK\cite{Nakatsuji2006}. We will return to the low energy spectrum in more detail at the end of this section. In their quantum chemistry calculations, Hozoi {\it et al.}\cite{Hozoi2014} showed that the energies of the {\it d-d} excitations are quite similar for a variety of pyrochlore iridates (R = Sm, Eu, Lu, and Y). In particular, the non-cubic contribution to the crystal field splitting is quite significant in all four materials, as demonstrated by the large splitting between the two peaks associated with intra-t$_{2g}$ transitions, $E_1 \approx 0.6$~eV and $E_2 \approx 0.9$~eV. The authors also determined that it is long-range lattice anisotropy, involving next-nearest neighbor Ir and R-site ions, which is primarily responsible for the symmetry-breaking field that results in the splitting of the $t_{2g}$ levels. Our RIXS measurements on $\rm Pr_2Ir_2O_7$ indicate that similar local {\it d-d} excitations are present even in this metallic sample, suggesting that the high-energy electronic structure of these materials may not be that different. A table of {\it d-d} excitation energies, extracted from fits to the RIXS data in Fig.~\ref{fig6}, is provided in Table 1.  We note that Pr$_2$Ir$_2$O$_7$ displays significantly larger non-cubic crystal field splitting than Eu$_2$Ir$_2$O$_7$ ($\Delta$ = 0.46 eV compared to 0.36 eV), but with a slight reduction in the octahedral crystal field splitting ($\sim$3.40 eV compared to 3.68 eV). 

\begin{table}
\caption{{\it d-d} excitation energies for Eu$_2$Ir$_2$O$_7$ and Pr$_2$Ir$_2$O$_7$ as determined from the RIXS data in Fig.~\ref{fig6}.  $E_1$ and $E_2$ correspond to transitions within the t$_{2g}$ manifold, while E$_3$ corresponds to transitions between the t$_{2g}$ and e$_{g}$ manifolds.  $\Delta$ represents the value of the non-cubic crystal electric field splitting.}
\centering 
\setlength{\tabcolsep}{8pt}
\begin{tabular}{c c c c c} 
\hline\hline 
\rule{0pt}{2.5ex} 
					            &	$E_1$	(eV) & $E_2$ (eV) & $E_3$ (eV) & $\Delta$ (eV) \\  
\hline 
\rule{0pt}{2.5ex}
Eu$_2$Ir$_2$O$_7$    	& 0.59(1)		& 0.95(1)		&	3.68(2)		& 0.36(2)  \\
Pr$_2$Ir$_2$O$_7$    	& 0.52(1)		& 0.98(1) 	& 3.40(2)		& 0.46(2)	 \\
\hline\hline 
\end{tabular}
\label{table1}
\end{table}

\begin{figure}
\begin{center}
\includegraphics[angle=0,width=3.1in]{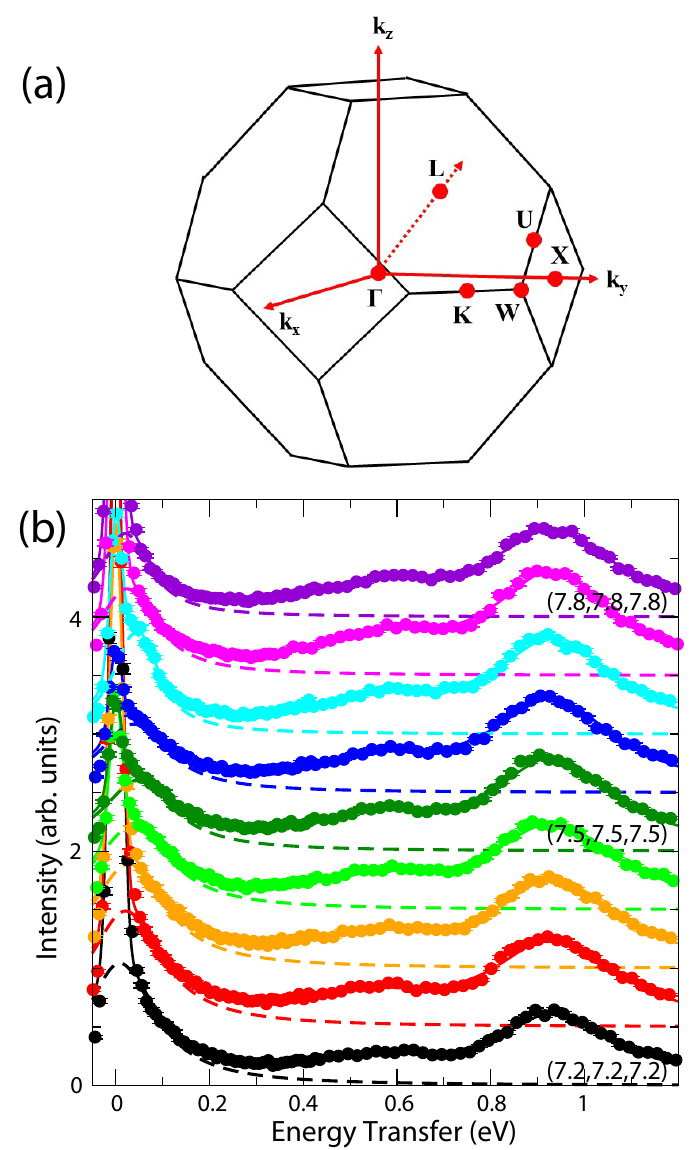}
\end{center}
\caption{(Color online) Momentum dependence of the Ir L$_3$-edge RIXS spectra for $\rm Eu_2Ir_2O_7$.  (a) A schematic drawing of the pyrochlore Brillouin zone.  The energy scans in (b) were collected along the $\Gamma-L-\Gamma$ direction marked by the dotted red line.  The low energy RIXS spectrum consists of two spin-orbit excitations at $\sim$0.6 and $\sim$0.9 eV (corresponding to transitions between the j$_{eff}$ = 1/2 and crystal-field split j$_{eff}$ = 3/2 levels), and a broad magnetic excitation below $\sim$0.15 eV.  All scans were collected at T = 25~K, well below the magnetic transition temperature for this material.}\label{fig7}
\end{figure}

In Fig.~\ref{fig7}, we plot the momentum dependence of the RIXS spectrum for $\rm Eu_2Ir_2O_7$.  These spectra were obtained for momentum transfers along the (111) direction in reciprocal space, tracing a $\Gamma-L-\Gamma$ path through the Brillouin zone as illustrated in Fig.~\ref{fig7}(a). The {\bf Q} points which correspond to the $\Gamma$ position, i.e. the zone centers at (7,7,7) and (8,8,8), were not measured in this study, since any spectra collected near allowed Bragg peak positions will be dominated by the strong elastic signal. The {\it d-d} excitations in $\rm Eu_2Ir_2O_7$ display no obvious momentum dependence, which is very different from square lattice iridates such as $\rm Sr_2IrO_4$\cite{JKim2012a,JKim2014} or $\rm Sr_3Ir_2O_7$ \cite{JKim2012b}. It should be emphasized that the strong momentum dependence in these square lattice iridates originates from the large hopping amplitude which contributes to strong superexchange coupling. Therefore, the bandwidth of the orbital (or spin-orbiton) excitations is the same as the magnon bandwidth. In the honeycomb iridate $\rm Na_2IrO_3$, smaller orbital overlap due to almost 90$^\circ$ Ir-O-Ir geometry keeps the dispersion of magnetic and orbital excitations small \cite{Gretarsson2013a,Gretarsson2013b}.

\begin{figure*}
\begin{center}
\includegraphics[angle=0,width=7in]{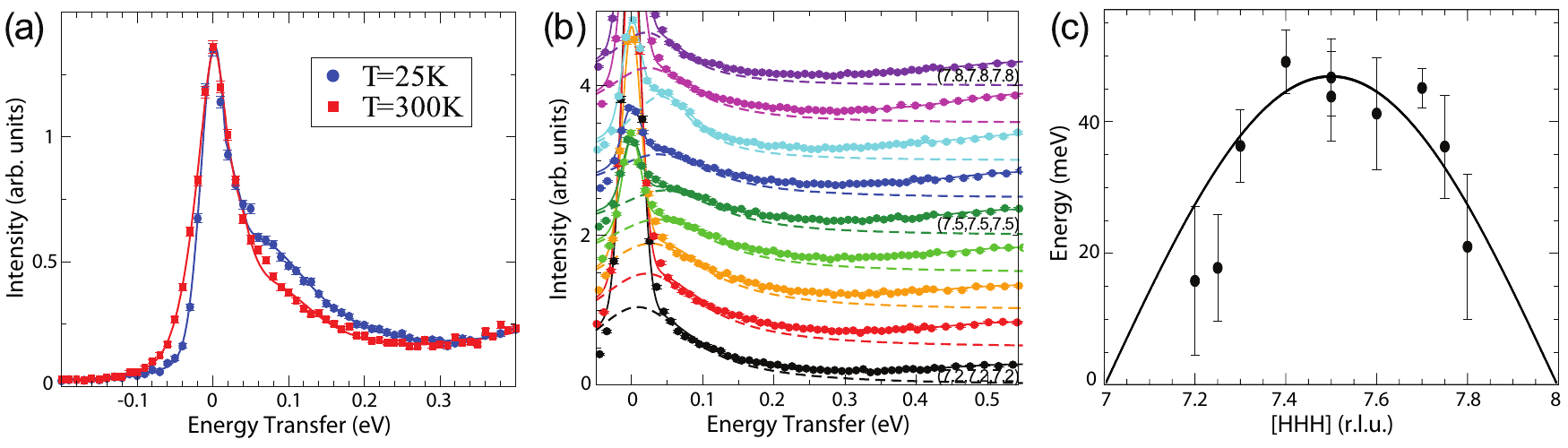}
\end{center}
\caption{(Color online) (a) Comparison of low energy RIXS spectra for $\rm Eu_2Ir_2O_7$ at {\bf Q} = (7.5, 7.5, 7.5) above (T = 300 K) and below (T = 25 K) the magnetic ordering transition.  (b) Momentum dependence of the low energy RIXS spectra at T = 25~K.  These scans represent the same data provided in Fig.~7(b), but with a focus on the broad, low-lying magnetic excitations.  (c) Dispersion of the magnetic excitation observed in (b).  The range of momentum transfers corresponds to a path along the $\langle$111$\rangle$ direction ($\Gamma-L-\Gamma$).  Details of the fitting procedure used to determine the magnetic peak positions are described in the main text.  
}\label{fig8}
\end{figure*}

The low energy shoulder below 0.3 eV does appear to display momentum dependence in Fig.~\ref{fig7}.  We focus on examining this low energy region in more detail in Fig.~\ref{fig8}.
In Fig.~\ref{fig8}(a), we show the temperature dependence of RIXS spectra collected at the {\bf Q} = (7.5, 7.5, 7.5) zone boundary position.  At T = 25~K, well below the magnetic transition temperature, we can clearly observe a broad feature below 0.1~eV. This feature is greatly suppressed when the temperature is raised above $T_M$, up to T = 300~K. This behavior can be contrasted with the negative energy transfer side of the spectrum (i.e. x-ray energy gain/sample energy loss), in which the tail of the elastic peak {\em increases} at high temperature. This form of temperature dependence can be explained in terms of an increase in the Bose population factor for phonons. However, the {\it decrease} of spectral weight observed at $\sim$0.1 eV requires an explanation that goes beyond routine temperature dependence. The most natural explanation is to associate this low-lying feature with magnetic excitations that lose spectral weight when magnetic order disappears at high temperature.

The low-lying magnetic excitation also shows significant momentum dependence. In Fig.~\ref{fig8}(b), we plot the low energy region of the spectra from Fig.~\ref{fig7}. Note that the shoulder on the elastic line appears to shift from higher energy to lower energy as {\bf Q} moves from the Brillouin zone boundary at (7.5, 7.5, 7.5) towards the zone centers at (7,7,7) and (8,8,8). This low-lying excitation is extremely broad, and therefore cannot be associated with a dispersive quasiparticle excitation. The simplest way to analyze the momentum dependence of this feature is to model the elastic line by a resolution-limited pseudo-Voigt function, and fit the position of the low energy shoulder using a simple Lorentzian function. This method provides a rough measure of the center of gravity of the spectral feature, but often overestimates the scattering intensity at negative energy transfers.  The quality of fit can be significantly improved by introducing an asymmetric lineshape, such as an exponentially damped Lorentzian, although the resulting peak positions do not appear to be overly sensitive to the precise choice of fit function.  

In Fig.~\ref{fig8}(c), we plot the dispersion of the low-lying magnetic excitation along the $\Gamma-L-\Gamma$ direction, using the peak positions extracted from our fitting analysis. Here the error bars have been chosen to reflect the additional uncertainty introduced by the choice of lineshape.  If we consider a simple spin wave model, the observed dispersion relation seems to suggest that the magnetic excitation reaches a maximum energy of 45 $\pm$ 5 meV near the zone boundary at (7.5, 7.5, 7.5) (i.e. the $L$-point), and disperses towards zero energy at the $\Gamma$ points on either side. This result would be consistent with a spin wave mode associated with {\bf q} = 0 magnetic order. However, it should be noted that the observation of such a highly damped excitation suggests that the simple spin wave picture is probably not appropriate for this material, and a more sophisticated interpretation of the spin dynamics is required.

\begin{figure}[b]
\begin{center}
\includegraphics[angle=0,width=3.1in]{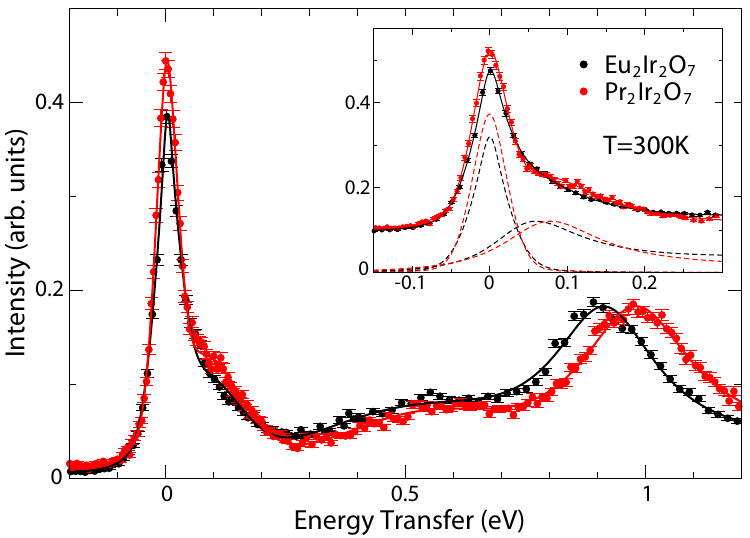}
\end{center}
\caption{(Color online) Comparison of RIXS spectra for Eu$_2$Ir$_2$O$_7$ and Pr$_2$Ir$_2$O$_7$.  Both scans were collected at T = 300 K and {\bf Q} = (7.5, 7.5, 7.5).  An enlarged view of the low energy RIXS spectra is provided in the inset.  The fit components that describe the elastic line and the lowest-lying excitation are illustrated by vertically offset dashed lines.}\label{fig9}
\end{figure}

In Fig.~\ref{fig9}, we compare the low energy RIXS spectra of Eu$_2$Ir$_2$O$_7$ and Pr$_2$Ir$_2$O$_7$.  Since Pr$_2$Ir$_2$O$_7$ displays no magnetic order down to T = 70 mK, here we have plotted two data sets collected at T = 300 K, well within the paramagnetic phase for both compounds.  As noted earlier, these measurements reveal a clear shift in the energy of the {\it d-d} excitations, with R = Pr displaying $\sim$25\% larger non-cubic crystal electric field splitting than R = Eu.  Most striking, however, is the fact that we observe a broad low energy shoulder in both of these materials, even outside the magnetically ordered state.  Modelling this shoulder with a damped Lorentzian lineshape, we find the center of the feature to be at 45 $\pm$ 5 meV for R = Eu and 70 $\pm$ 5 meV for R = Pr.  Interpreting the origins of this feature is non-trivial, as low-lying spectral weight can arise from phonon scattering, electon-hole continuum scattering, and/or paramagnetic fluctuations above $T_M$.  However, the fact that this shoulder occurs at substantially higher energies for R = Pr would appear to contradict both the resonant phonon explanation (since the difference in phonon energies for isostructural Eu$_2$Ir$_2$O$_7$ and Pr$_2$Ir$_2$O$_7$ would be expected to be much smaller) and the electron-hole continuum explanation (since R = Pr is more metallic than R = Eu, we would expect the continuum scattering in Pr$_2$Ir$_2$O$_7$ to occur at lower energies rather than higher ones).  Similarly, the fact that the high temperature shoulder in Eu$_2$Ir$_2$O$_7$ appears at exactly the same energy as the low temperature magnetic excitation, albeit with significantly reduced spectral weight, strongly suggests that this is a paramagnetic remnant of the same magnetic feature.  Therefore, we tentatively attribute this low-lying feature to the presence of short-lived paramagnetic fluctuations.

\section{Discussion and conclusions}

\subsection{Structure}

The main conclusion from our structural studies is that the pyrochlore iridate Eu$_2$Ir$_2$O$_7$ retains its $Fd\bar{3}m$ crystal symmetry throughout a wide variety of electronic and magnetic phases. We find no evidence of any structural distortion associated with the thermally-driven magnetic transition at $T_M$, confirming previous neutron and x-ray diffraction results. In addition, we observe no pressure-induced structural transitions or symmetry changes for applied pressures up to 17 GPa. This result has direct implications for recent high pressure transport studies by Tafti {\it et al.}\cite{Tafti2012}, which reported pressure-induced metal-insulator (low temperature) or metal-incoherent metal (high temperature) transitions in Eu$_2$Ir$_2$O$_7$ around 6-8 GPa. Since our experimental results have demonstrated that no structural anomalies occur in the vicinity of the critical pressure, we can conclude that these pressure-induced transitions are likely driven by purely electronic effects.

Our resonant x-ray scattering measurements provided evidence of temperature-independent ATS scattering, but were unable to detect magnetic Bragg peaks. This differs from previous resonant x-ray scattering measurements on single crystal $\rm Eu_2Ir_2O_7$ by Sagayama {\it et al.}, which reported a resonantly enhanced peak at (10,0,0) that disappeared above the magnetic transition temperature. We note that (10,0,0) is a forbidden peak position which is crystallographically equivalent to the (4,4,2) position measured in this study, with both wave vectors corresponding to an {\it X}-point on the Brillouin zone boundary (see Fig.~\ref{fig7}(a)). It is possible that this discrepancy is simply due to differences in sample stoichiometry, as the single crystal sample used in our study was found to be Eu-rich (x = -0.09) while the sample in Ref. 24 was found to be slightly Ir-rich (x $<$ 0.05).  However, bulk magnetization measurements on our sample do reveal a clear magnetic transition at $T_M$ $\sim$ 155 K, and in spite of the lack of magnetic Bragg peaks below $T_M$, there is strong evidence of magnetic dynamics provided by RIXS. This implies that at the very least there must be some form of short-range magnetic correlations in this sample.

\subsection{Excitations}

The main result from our RIXS investigation of $\rm Eu_2Ir_2O_7$ is that there is a highly damped magnetic excitation which reaches a maximum energy of 45 $\pm$ 5 meV at the $L$-position.  This excitation exhibits significant momentum dependence, and disperses towards lower energies as the momentum transfer approaches the $\Gamma$-position. The observation of such Landau-damped magnetic excitations is not surprising given that our sample remains metallic/semi-metallic even at low temperatures due to sample stoichiometry. As discussed in Section II and in previous work by Ishikawa {\it et al.}\cite{Ishikawa2012}, many $\rm Eu_2Ir_2O_7$ single crystal samples are known to suffer from (1) deviations from ideal stoichiometry, and (2) anti-site disorder which leads to mixing of the Eu and Ir sublattices. It is unclear what kind of magnetic excitations should be expected in such an impurity/defect-induced metallic state.

\begin{figure}[t]
\begin{center}
\includegraphics[angle=0,width=3.1in]{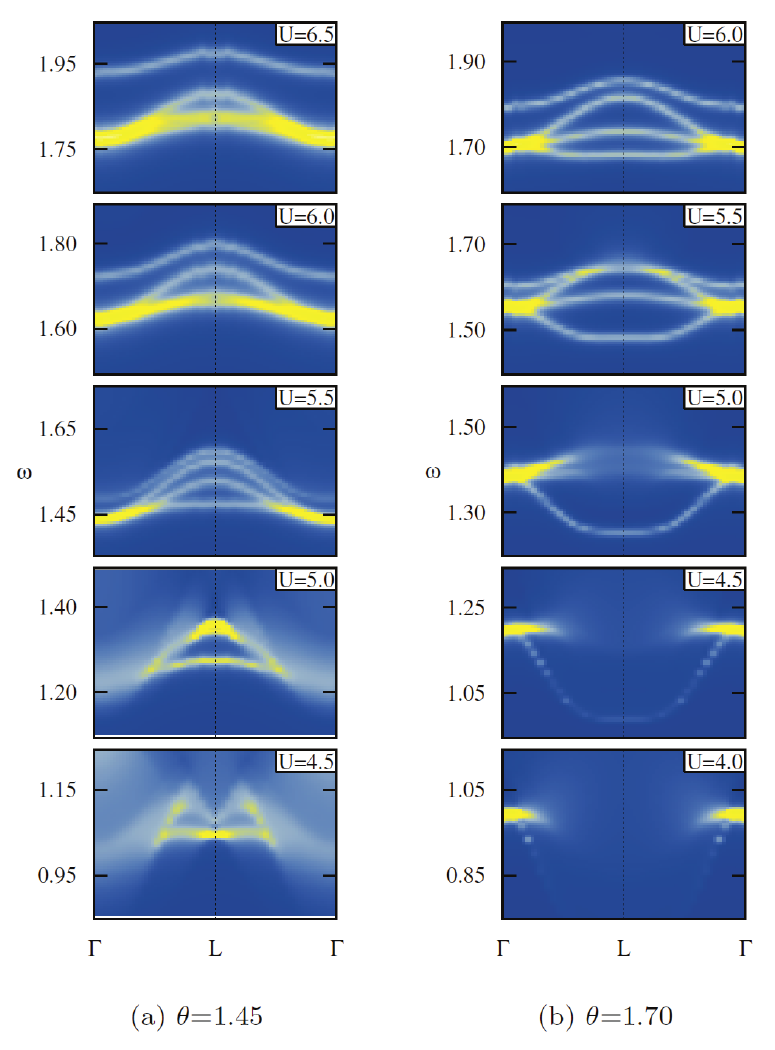}
\end{center}
\caption{(Color online) RPA dynamical structure factors for (a) $\theta = 1.45$ and (b) $\theta = 1.70$ for various values of U, adapted from Ref.~\onlinecite{EKHLee2013}.  The value of $\theta$ parameterizes hopping amplitudes as described in Ref.~\onlinecite{EKHLee2013}.  These two values roughly correspond to parameters having (a) (topological) semi-metallic and (b) metallic ground states, respectively.  Sharp dispersion can be seen at larger values of U, while lighter intensities and broadened spectra are observed at lower U due to Landau damping. }\label{fig10}
\end{figure}

Another way to understand the damped magnetic excitations observed in our sample is to approach the problem using weak/intermediate coupling theory. Since the ground state of Eu$_2$Ir$_2$O$_7$ is close to a metal, conventional semi-metal, or Weyl semi-metal phase, it is reasonable to assume that well-defined magnon modes derived from a strong-coupling approach will be inadequate to describe the magnetic excitations in this material. In recent theoretical studies by Lee {\it et al.}\cite{EKHLee2013}, the dynamic structure factor for the pyrochlore iridates, $S({\bf q},\omega)$, was calculated using the random phase approximation (RPA). This result is reproduced in Fig.~\ref{fig10} for the same $\Gamma-L-\Gamma$ path measured in Figs. 7 and 8.  Note that this calculation has been performed for two different regimes from the theoretical phase diagram (See Ref. \onlinecite{EKHLee2013}), with hopping values correponding to (a) a (topological) semi-metallic phase ($\theta$ = 1.45), and (b) a metallic phase ($\theta$ = 1.70). As expected, in the strong coupling limit (U = 6), dispersive quasiparticle excitations are found. However, as the coupling strength decreases, the overall spectrum begins to broaden. In the intermediate coupling limit (U = 4 to 4.5), the sharp quasiparticle features have almost vanished, and the only well-defined spectral features are found near special high symmetry positions.

What is interesting is that these sharp high symmetry points differ greatly between the two regions of the phase diagram considered in Ref. \onlinecite{EKHLee2013}. The $S({\bf q},\omega)$ in Fig.~\ref{fig10}(a), obtained with parameters close to the Weyl semi-metal phase, has a well-defined peak near the $L$-point. On the other hand, the $S({\bf q},\omega)$ in Fig.~\ref{fig10}(b), which corresponds to the metallic all-in/all-out phase, displays a strong, sharp peak near the $\Gamma$-point. Our RIXS spectrum exhibits a strong peak at the $L$-point, which broadens and weakens as it approaches the $\Gamma$-point.  Thus, the experimental data clearly supports the calculated $S({\bf q},\omega)$ shown in Fig.~\ref{fig10}(a).

One important caveat is whether it is justified to compare the RIXS spectrum with $S({\bf q},\omega)$ in this situation. In the case of localized spin models, this certainly appears to be a valid approach. In their study of $\rm Sr_2IrO_4$, Kim {\it et al.} showed that the magnetic RIXS spectrum is proportional to the dynamic structure factor for magnon excitations. Similarly, the RIXS spectrum of $\rm La_2CuO_4$ has also been found to follow the dynamic stucture factor\cite{Braicovich2009}. In systems where electrons take on more itinerant characteristics, there appears to be more controversy on this issue, and it is clear that further studies are still needed. One comment that we can offer here is that possible intensity modulations due to matrix-element effects, which are ignored in the $S({\bf q},\omega)$ calculation, are not significant in our measurements. These transition matrix elements depend on the total photon momentum transfer ${\bf Q}$, and not the crystal momentum transfer ${\bf q} \equiv {\bf Q} - {\bf G}$. The total momentum transfer along our $\Gamma-L-\Gamma$ line ranges from (7,7,7) to (8,8,8), i.e. only $\sim$13\% change in ${\bf Q}$.

Perhaps the most surprising result from our RIXS investigation concerns the excitation spectrum of Pr$_2$Ir$_2$O$_7$.  In spite of the absence of magnetic order in this compound, we observe a similar highly damped $L$-point feature at 70 $\pm$ 5 meV. This feature closely resembles the low energy peak in {\it paramagnetic} Eu$_2$Ir$_2$O$_7$, and thus we have tentatively attributed it to paramagnetic fluctuations.  The R = Pr sample exhibits a low-temperature anomalous Hall effect, which has been associated with an exotic metallic spin liquid state\cite{Machida2007,Machida2010}. It has been proposed that the large Pr$^{3+}$ moments exhibit spin-ice-like physics, and that strong quantum fluctuations drive ``quantum melting'' into a chiral spin liquid\cite{Onoda2010,Machida2010}.  The Pr moments are indirectly coupled to the itinerant Ir electrons through an antiferromagnetic RKKY interaction.  The observed Curie-Weiss temperature of $\theta_{CW}$ = 20 K, which is much larger than the energy scale expected from the Pr moments, is attributed to this RKKY energy scale\cite{Nakatsuji2006}.  Below T $\sim$ 20 K, the Pr moments are partially screened by the Kondo effect, and the antiferromagnetic interaction is renormalized to $\theta_{CW}$ $\sim$ 1.7 K\cite{Nakatsuji2006}.  Our observation of resonant inelastic scattering intensity at 70 meV $\approx$ 810 K suggests that the energy scale associated with the Ir conduction electrons is much greater than either of these interactions. Therefore, our results point towards the existence of a third, significantly larger, magnetic energy scale in Pr$_2$Ir$_2$O$_7$.

\begin{acknowledgments}
We would like to thank Arun Paramekanti and Jacob Ruff for valuable discussions. Work at the University of Toronto was supported by the Natural Sciences and Engineering Research Council of Canada through a Discovery Grant and Research Tools and Instruments Grant.  Work at Seoul National University was supported by National Creative Research Initiative 2010-0018300. Use of the Advanced Photon Source at Argonne National Laboratory is supported by the U.S. Department of Energy, Office of Science, under Contract No. DE-AC02-06CH11357.  Use of the Canadian Light Source is supported by the Canada Foundation for Innovation, the Natural Sciences and Engineering Research Council of Canada, the University of Saskatchewan, the Government of Saskatchewan, Western Economic Diversification Canada, the National Research Council of Canada, and the Canadian Institutes of Health Research.
\end{acknowledgments}

\bibliography{227_refs}

\newpage
\appendix
\section{Sample Characterization (Supplemental Material)}

Previous experimental studies of pyrochlore iridates have demonstrated that sample stoichiometry plays an important role in determining electronic properties\cite{Ishikawa2012}.  It is very difficult to synthesize stoichiometric single crystal samples of Eu$_2$Ir$_2$O$_7$, and the resulting material is more accurately described as Eu$_{2(1-x)}$Ir$_{2(1+x)}$O$_{7+\delta}$.  Even a $\sim$1\% deviation from ideal stoichiometry is sufficient to change the residual resistivity ratio of a sample by several orders of magnitude\cite{Ishikawa2012}.  Samples near ideal stoichiometry display a clear metal-insulator transition at $T_C$, with $\rho(2K)\/\rho(300K)$ on the order of 10$^3$ to 10$^5$.  In contrast, samples which display a significant excess of Eu or Ir tend to remain metallic or semi-metallic at low temperature.  Sample stoichiometry has comparatively little influence on the magnetic properties of Eu$_2$Ir$_2$O$_7$.  For x $<$ 0.03, non-stoichiometry reduces the hysteresis between field-cooled and zero-field-cooled magnetic susceptibility, but has no apparent effect on the value of $T_C$\cite{Ishikawa2012}.  

As noted in the main text, EPMA measurements on the single crystal used in this study suggest that our sample is Eu-rich/Ir-deficient, with a composition of x = -0.09(2) and $\delta$ = 0.06(2).  Bulk characterization measurements on this sample are provided in Figure 11.  Note that for this Eu-rich stoichiometry we observe a clear splitting of the magnetic and electronic transition temperatures.  The magnetic transition occurs at $T_M$ $\sim$ 155 K, with pronounced field-cooled/zero-field-cooled hysteresis appearing below 140 K.  Although our sample is not strongly insulating at low temperatures ($\rho_{4K}$/$\rho_{300K}$ $\sim$0.7), the temperature dependence of the resistivity shows a clear upturn below $T_E$ $\sim$ 60 K and an inflection point at $\sim$130 K.  This behavior can be contrasted with the simultaneous magnetic ordering and metal-insulator transitions that occur in ``near-stoichiometric'' samples at $T_C$ = 120 K.

\begin{figure}
\begin{center}
\includegraphics[angle=0,width=3.1in]{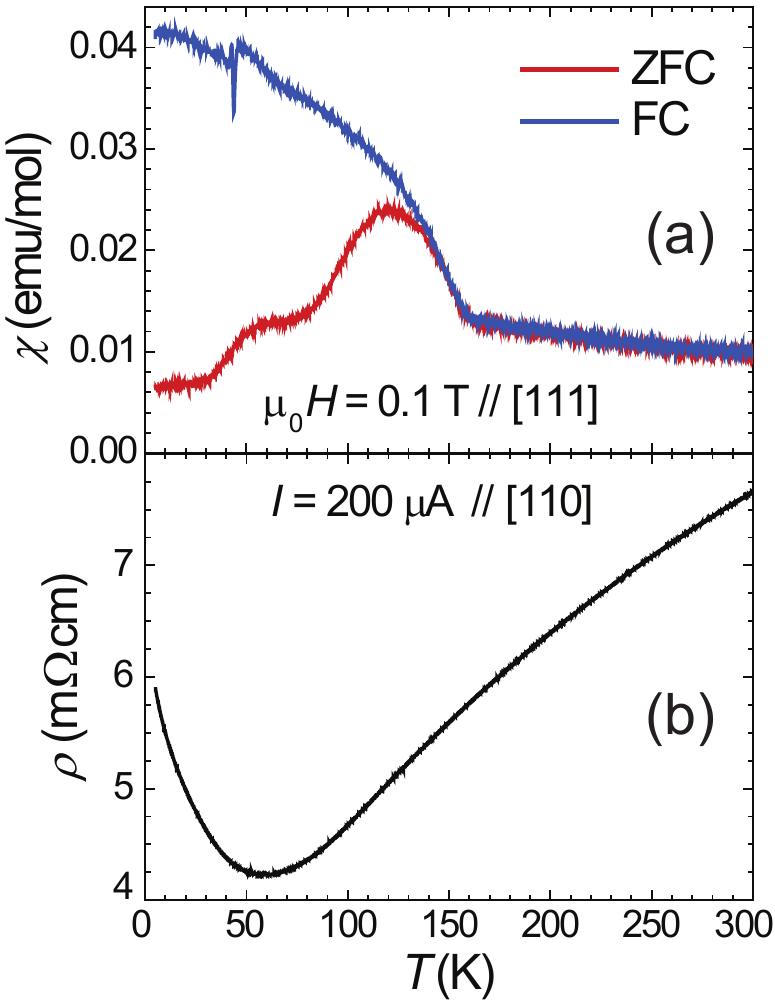}
\end{center}
\caption{(Color online) Bulk characterization measurements on single crystal Eu$_{2.18}$Ir$_{1.82}$O$_{7.06}$ showing the temperature dependence of (a) magnetic susceptibility and (b) electrical resistivity. }\label{sfig1}
\end{figure}

\section{High Pressure Structural Refinements (Supplemental Material)}

The pressure dependence of the crystal structure of Eu$_2$Ir$_2$O$_7$ was investigated using high pressure x-ray powder diffraction measurements. Diffraction data was collected using a two-dimensional image plate detector, and processed using the FIT2D program\cite{Fit2D}.  The resulting one-dimensional diffraction patterns (Intensity vs. 2$\theta$) were then refined using the GSAS software package\cite{GSAS}.  The background, cubic lattice parameter ({\it a}), lineshape, oxygen positional parameter (X(O1)), and thermal parameters (U$_{iso}$) were all refined individually, and then simultaneously.  A series of representative refinements are provided in Figure 12.  These refinements were carried out on data sets collected at pressures of 0.1, 6.9, 11.0, and 14.5 GPa, using silicone fluid as a pressure transmitting medium.  The key structural parameters from these refinements are provided in Table II.

\begin{figure}
\begin{center}
\includegraphics[angle=0,width=3.1in]{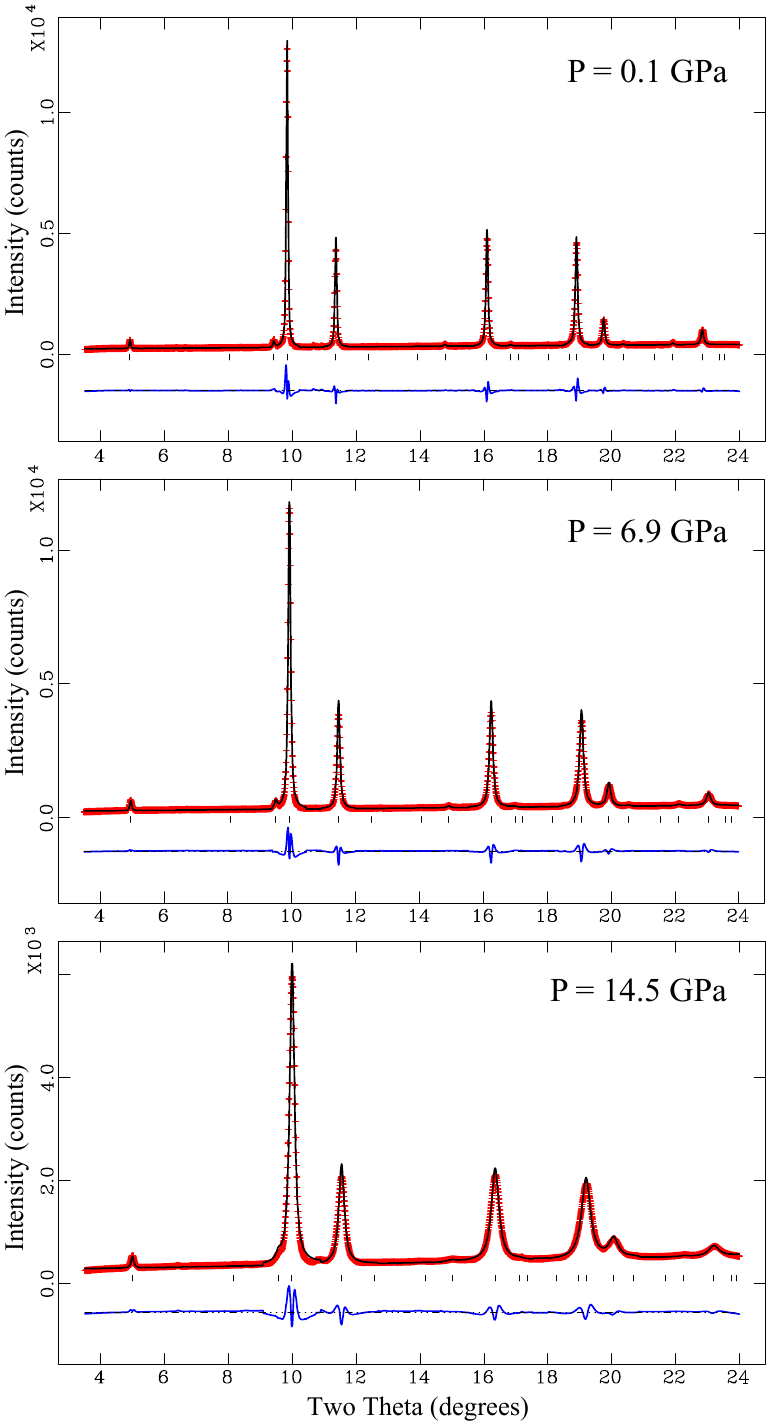}
\end{center}
\caption{(Color online) Representative Rietveld refinements at P = 0.1, 6.9, 11.0, and 14.5 GPa.  Data points are marked by red crosses, the Rietveld fit is marked by the solid black line, and the difference between fit and data is marked by the solid blue line. }\label{sfig2}
\end{figure}

\begin{table}[b]
\caption{Refined structural parameters for Eu$_2$Ir$_2$O$_7$ at several representative pressures.}
\centering 
\setlength{\tabcolsep}{6pt}
\begin{tabular}{c c c c c} 
\hline\hline 
\rule{0pt}{2.5ex} 
					& 0.1 GPa		& 6.9 GPa 	& 11.0 GPa 	& 14.5 GPa \\  
\hline 
\rule{0pt}{2.5ex}
a ({\AA})	& 10.2880(1) 	& 10.1915(2)	&	10.1566(3)	& 10.1287(4)	\\
x(O1)    	& 0.3290(13)		& 0.3288(13) 		& 0.3306(16)		& 0.3341(16)	\\
\hline\hline 
\end{tabular}
\label{table2}
\end{table}

Given concerns with stoichiometry in single crystal Eu$_2$Ir$_2$O$_7$, we also refined the occupancies of the Eu and Ir sites in our powder data.  Negligible improvement in goodness-of-fit was achieved by (a) varying the Eu and Ir occupancies freely or (b) refining the occupancies under a ``fully occupied'' constraint (i.e. x$_{Eu}$ + x$_{Ir}$ = 2).  At ambient pressure, these refinements yielded occupancies of x$_{Eu}$ = 0.985(6) and x$_{Ir}$ = 1.015(6). Thus, we can conclude that the powder sample is much closer to ideal stoichiometry than the single crystal sample, and that it is slightly Ir-rich rather than Ir-deficient.  

The diffraction patterns in Figure 12 reveal a clear broadening of the lineshape at higher pressures.  This is the result of pressure-induced strain broadening, which can be modelled by refining lineshape parameters during the structural refinements.  In particular, the Lorentzian component of the lineshape profile, Ly, provides a quantitative measure of this microstrain broadening.  The pressure dependence of the Ly parameter is plotted in Figure 13(a).  Note that a sudden increase in peak broadening is observed for both pressure transmitting media, occurring at $\sim$8 GPa for MEW 16:5:1 and $\sim$10 GPa for silicone fluid.  This jump likely provides an indication of where hydrostatic conditions begin to break down in each of these data sets.

The goodness-of-fit parameters, R$_{wp}$ and R$_{p}$, are plotted in Figure 13(b).  Note that the goodness-of-fit remains relatively constant over the entire pressure range, and even improves slightly at higher pressures.  This provides a strong indication that the $Fd\bar{3}m$ structural model remains valid for Eu$_2$Ir$_2$O$_7$ up to the highest measured pressures.  If there were any pressure-induced structural distortions or symmetry changes in this material we would expect this to be reflected by a decrease in the quality of the refinements. 

\begin{figure}[b]
\begin{center}
\includegraphics[angle=0,width=3.1in]{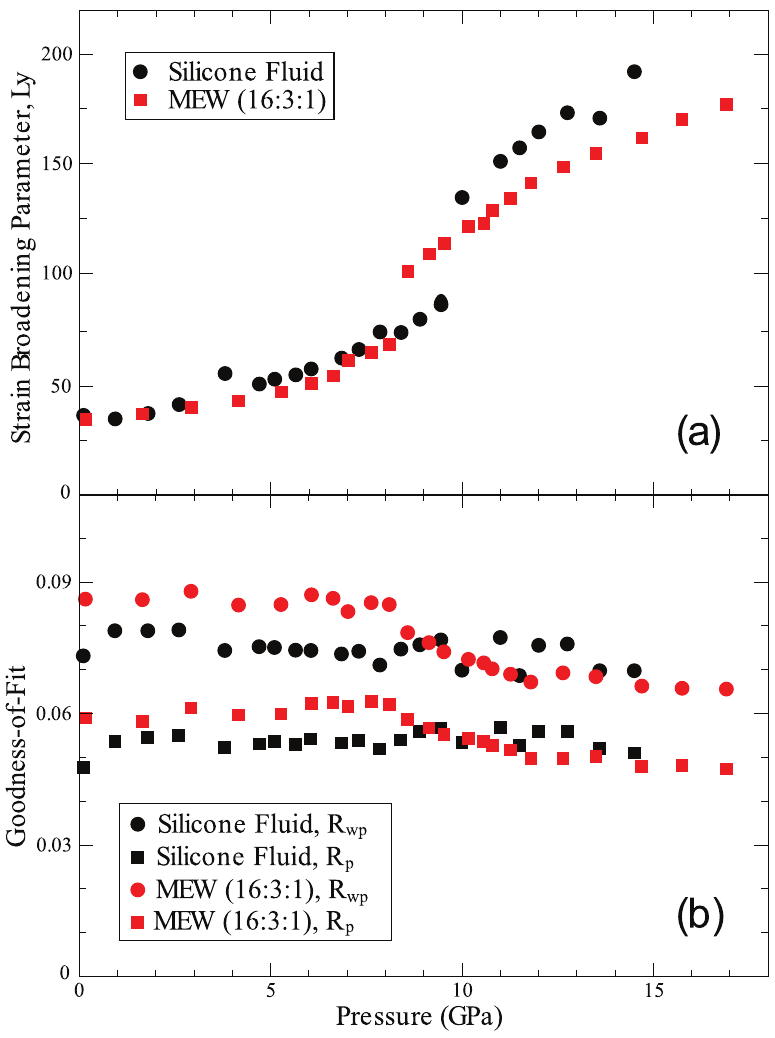}
\end{center}
\caption{(Color online) The pressure dependence of (a) the lineshape parameter Ly and (b) the goodness-of-fit parameters R$_{wp}$ and R$_p$.  Ly describes the effect of strain broadening on the diffraction pattern, which increases as a function of pressure.  The goodness-of-fit parameters indicate that the $Fd\bar{3}m$ structural model remains valid up to P $\sim$ 17 GPa.}\label{sfig3}
\end{figure}

\end{document}